\documentclass[conference]{IEEEtran}
\IEEEoverridecommandlockouts

\usepackage{cite}
\usepackage{amsmath,amssymb,amsfonts}
\usepackage{algorithmic}
\usepackage{graphicx}
\usepackage{textcomp}
\usepackage{xcolor}
\usepackage{url}
\usepackage[nolist]{acronym}
\usepackage[utf8]{inputenc}
\def\BibTeX{{\rm B\kern-.05em{\sc i\kern-.025em b}\kern-.08em
    T\kern-.1667em\lower.7ex\hbox{E}\kern-.125emX}}

\def\code#1{\texttt{#1}}
\begin{document}

\title{Integrating Deterministic Networking with 5G \vspace{-0.8cm}\thanks{This work received funding from Bavarian State Ministry for Economic Affairs, Regional Development and Energy (StMWi) project KI.FABRIK (grant no. DIK0249) and "6G Futre Lab Bavaria", Federal Ministry of Education and Research of Germany (BMBF) project 6G-life (16KISK002) and Deutsche Forschungsgemeinschaft (DFG, German Research Foundation) - 316878574..
\vspace{-0.5cm}}}


\newcommand{\phil}[1]{\textcolor{red}{[P: #1]}}
\newcommand{\yash}[1]{\textcolor{purple}{[Y: #1]}}

\author{\IEEEauthorblockN{Yash Deshpande\IEEEauthorrefmark{1}, Philip Diederich\IEEEauthorrefmark{1}, Muhamad Luthfi\IEEEauthorrefmark{1}, Laura Becker\IEEEauthorrefmark{1}, José Fontalvo-Hernández\IEEEauthorrefmark{2}, Wolfgang Kellerer\IEEEauthorrefmark{1}\\
\IEEEauthorblockA{\IEEEauthorrefmark{1}Chair of Communication Networks, Technical University of Munich, Germany \\
\IEEEauthorrefmark{2}Siemens AG,  Germany\\
Email: \{yash.deshpande,philip.diederich, muhamad.luthfi, laura.alexandra.becker, wolfgang.kellerer\}@tum.de}, jose-eduardo.fontalvo-hernandez@siemens.com}
\vspace{-1cm}}

\maketitle

\begin{abstract}
The rising prevalence of real-time applications that require deterministic communication over mobile networks necessitates the joint operation of both mobile and fixed network components.
This joint operation requires designing components that interact between the two technologies to provide users with latency and packet loss guarantees.
In this work, we demonstrate a fully integrated 5G-DetNet that can guarantee the end-to-end demands of different flows. 
Moreover, we show how such a network can be implemented using low-cost hardware and open-source software, making it accessible to many 5G testbeds.  
The features demonstrated in this work are a network manager that does the routing and scheduling, an application function in the 5G core that interfaces with the network manager, and a network-side translator for user-plane management and de-jittering of the real-time streams. 
\end{abstract}

\begin{acronym}
    \acro{LCDN}{Low-Cost Deterministic Networking}
    \acro{AP}{access point}
    \acro{PTP}[PTP]{Precision Time Protocol}
    \acro{OWD}{One-way delay}
    \acro{RTT}{Round-trip time}
    \acro{SMS}{Smart manufacturing systems}
    \acro{CPS}{Cyber-physical systems}
    \acro{IIoT}{Industrial Internet-of-Things}
    \acro{TUB}{Time Uncertainty Bound}
    \acro{UTC}{Coordinated Universal Time}
    \acro{GPS}{Global Positioning System}
    \acro{OWD}{One-way delay}
    \acro{ppm}{parts per million}
    \acro{RTT}{Round trip time}
    \acro{NIC}{Network interface controllers}
    \acro{DES}{Discrete event simulator}
    \acro{MAC}{media access control}
    \acro{TSN}{Time Sensitive Networking}
    \acro{NTP}{Network Time Protocol}
    \acro{COTS}{commercial off-the-shelf}
    \acro{PDV}{packet delay variation}
    \acro{LAN}{local area networks}
    \acro{SNMP}{Simple Network Management Protocol}
    \acro{SDN}{Software defined networking}
    \acro{NIC}{Network Interface Card}
    \acro{NC}{Network Calculus}
    \acro{MDR}{Maximum Drift Rate}
    \acro{TDD}{Time Division Duplex}
    \acro{NR}{New Radio}
    \acro{PHC}{PTP Hardware Clock}
    \acro{TT}{TimeTether}
    \acro{API}{Application Programming Interface}
    \acro{DNC}{Deterministic Network Calculus}
    \acro{SP}{strict-priority}
    \acro{IWRR}{Interleaved Weighted Round-Robin}
    \acro{RB}{resource blocks}
    \acro{UE}{User Equipment}
    \acro{OFDM}{Orthogonal Frequency Division Multiplexing}
    \acro{SRS}{Sounding Reference Signal}
    \acro{UL}{Uplink}
    \acro{RRC}{Radio Resource Control}
    \acro{IoT}{Internet of Things}
    \acro{QoS}{Quality of Service}
    \acro{LCDN}{Low-Cost Deterministic Networking}
    \acro{E2E}{End-to-End}
    \acro{JRS}{Joint Routing and Scheduling}
    \acro{TAS}{Time-Aware Shaper}
    \acro{CBS}{Credit-Based Shaper}
    \acro{SPQ}{Strict Priority}
    \acro{ATS}{Asynchronous Traffic Shaper}
    \acro{TC}{Traffic Control}
    \acro{ILP}{Integer Liner Programming}
    \acro{IWRR}{Interleaved Weighted Round Robin}
    \acro{SRP}{Stream Reservation Protocol}
    \acro{RAP}{Resource Allocation Protocol}
    \acro{SOTA}{state-of-the-art}
    \acro{USRP}{Universal Software Radio Peripheral}
    \acro{RAN}{Radio access network}
    \acro{SMF}{Session Management Function}
    \acro{UPF}{User Plane Function}
    \acro{AF}{Application Function}
    \acro{CNM}{Central Network Manager}
    \acro{MCS}{Modulation and Coding Scheme}
    \acro{NW-TT}{Network-side TSN translator}
    \acro{5GS}{5G system}
    \acro{TBS}{Transport Block Size}
    \acro{LCDN}{Low-Cost Deterministic Network}
    \acro{CCPF}{Centralized Controller Plane function}
\end{acronym}
\section{Introduction}
\label{sec:intro}
Emerging applications such as \ac{IoT}, teleoperation, and next-generation manufacturing have heightened the demand for time-sensitive and reliable network traffic. 
These applications necessitate precise \ac{E2E} latency and packet delivery guarantees to function effectively. 
While traditional networks prioritize network stability and reliability, they often fail to ensure the real-time data delivery required by the above-listed modern applications. 
To address this gap, IEEE 802.1 \ac{TSN} and IETF DetNet provide deterministic connectivity and Quality of Service (QoS) within Ethernet and IP networks.

Most emerging time-sensitive applications require wireless communication because of mobility or infrastructure constraints. 
Due to wide-area coverage, bespoke network management including Quality of Service, and high data rates, 5G and emerging 6G networks are promising candidates to fulfill the wireless requirements of these applications.   
Integrating TSN or DetNet with 5G is crucial to extend wired networks' deterministic capabilities and provide \ac{E2E} determinism across networks~\cite{tsn5gintegratoin}. 

The 5G and TSN/DetNet convergence has been introduced in 3GPP Release 16~\cite{3gpp2020} and 18. 
This standard recommends adding a device-side translator, a network-side translator, and an \ac{AF} in the mobile network core to connect a 5G network to a DetNet. 
The first two should be used for the user plane, and the last one for control plane integration of DetNet and 5G. 
Then, the whole 5G system can be seen as a \textit{transit node} or a forwarding device in the larger DetNet network. 
Most DetNet networks require their nodes to have deterministic features. 
The per-class or per-flow delay and buffer size should be known and bounded. 
However, a 5G system is complex; therefore, it is difficult to determine these quantities for a 5G system and report them to a DetNet controller \cite{bridgeDelay}.
Parameters such as bandwidth, scheduling algorithm, numerology, and TDD patterns can impact the packet latencies in a 5G system. 
Furthermore, conditions such as channel quality, network load, and the number of connected users can lead to a dynamic change of the real-time behavior of a 5G system.  
It is crucial for a typical DetNet device to work in close coordination with other bridges and endpoints in the network. Therefore, realizing the 5G system interfaces - even according to the prescribed standard - with the DetNet system requires special attention and bespoke implementation.   

Experimental validation and evaluation of emerging technologies on hardware platforms are pivotal for their acceptance within the industry. 
Complex systems such as 5G and DetNet need to be rigorously validated and evaluated by academia on research platforms and testbeds. 
This ensures reliable performance before such systems must be produced and deployed at scale.
Our work in this paper outlines integrating a low-cost DetNet network with a 5G system to demonstrate \ac{E2E} packet delivery guarantees on a 5G testbed. 
To our knowledge, such a combined system - built according to 3GPP recommendations - has been previously evaluated on simulations~\cite{5GTQ,BridgeSim,MultiStreamTSN5G} omitting hardware-level implementation that is presented in this work.
The hardware implementations in existing literature~\cite{integrationFRER,aijaz2024time}, do not demonstrate an \ac{E2E} solution, with a full DetNet and 5G, but focus only on the \ac{RAN} part of 5G. 
Next to providing the design of our demonstrator setup using low-cost compoenents, the lessons learned from our demonstration are helpful for the wider academic community working on problems related to 5G-DetNet integration.

\begin{figure}
    \centering
    \includegraphics[width=\columnwidth]{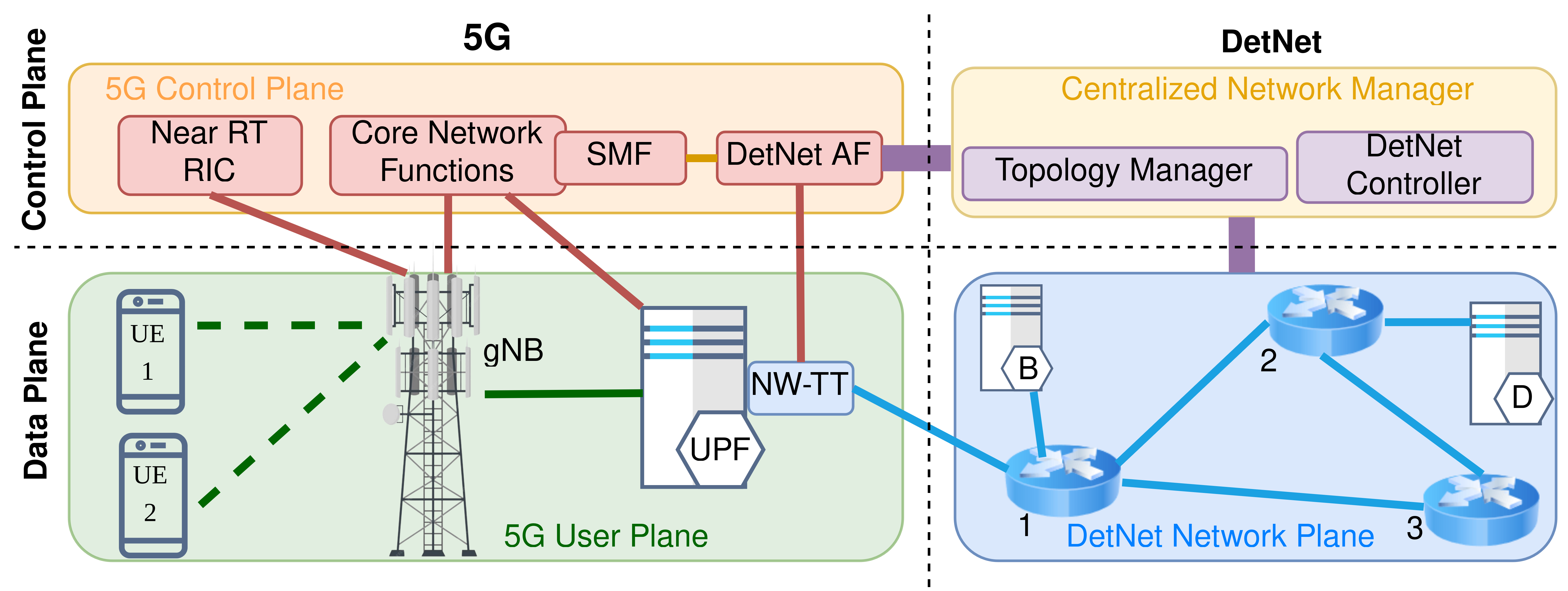}
    \vspace{-0.8cm}
    \caption{Overview of the demonstration testbed. The left half is the 5G system, the right half is the DetNet system, the bottom half is the data plane, and the top half is the control plane. Two UEs produce traffic with the same destination (D) in the DetNet. DetNet's CNM registers and configures the flows to meet their packet delay deadlines. The CNM connects to the DetNet AF in the 5G control plane to do this. }
    \label{fig:setup}
\end{figure}

\section{System Design}
\label{sec:system_desing}

\subsection{DetNet System}
\label{subsec:detnet}
The DetNet system is shown in the right part of Figure~\ref{fig:setup}. 
We use \ac{LCDN}~\cite{LCDN} as the fixed DetNet system derived from Chameleon\cite{chameleon}.
\ac{LCDN} is a DetNet method that can be implemented on low-cost consumer-grade switches and end-hosts. 
It performs joint routing and scheduling to organize flows in a network with low-cost switches to meet the latency and packet loss guarantees specified for the flow. 
LCDN uses source routing and policing of flows at the end hosts.
The \ac{CNM} in \ac{LCDN} is the DetNet \ac{CCPF} that decides all the schedules and routes.  
Regardless of the DetNet method used, the function of any \ac{CCPF} is to (1) keep track of the network topology, (2) keep track of the flows present in the network, (3) set up the end hosts and switches to perform joint routing and scheduling, (4) accept, reject or reconfigure flows in the network. 
The \ac{CNM} needs to know specific properties of the switch, such as the forwarding delay per class and buffer sizes.
LCDN obtains the \code{lldp} information to discover and keep track of the topology and a lightweight Linux middleware in the end-hosts that can be configured via HTTP. 
We use the same switches as in the LCDN~\cite{LCDN} paper; hence, these properties for the switches are apriori known. 

Before starting a flow, the end host requests the \ac{CNM} (via the middleware) to determine whether a flow can be installed in the network.  
This request contains the flow's traffic specification (TSpec) - the source and destination IP, the burst and rate of the flow, and the required latency bound.

The \ac{CNM} checks the topology, performs the \ac{LCDN} optimization pipeline, and returns the path of the flow (VLAN ID), the class of flow (VLAN priority). 
Henceforth, the middleware tags every packet for this flow with the given VLAN and VLAN priority and runs a policing function that maintains that flow's rate and burst specifications.
This tagging procedure in the middleware is performed by the Linux \code{tc} utility and does not require any custom-built packages. 
A RESTful server in the middleware communicates with the \ac{CNM}.

\subsection{5G System}
\label{subsec:5G_system}
The general architecture of the 5G system in the demo is shown in the left half of Figure \ref{fig:setup}.
The 5G system should also look like a \textit{transit node} and report its parameters to the DetNet controller~\cite{3gpp2020}.
However, these parameters for a 5G system are more difficult to ascertain.
First, since no \code{lldp} protocol runs in the 5G system, the UEs connected to it must be reported to the \ac{CNM} using a subscription to \ac{SMF} application. 
We implemented a DetNet \ac{AF} that connects to the DetNet controller via an HTTP interface.
The \ac{AF} subscribes to the \ac{SMF} and constantly polls to check the currently active UEs and reports them to the \ac{CNM}.
Moreover, the \ac{AF} subscribes to the indication messages of the gNB via the Near-RT RIC to know the current \ac{TDD} pattern and scheduler policy as well as the channel quality and \ac{MCS} of a UE. 
Then, the \ac{AF} calculates the worst-case uplink and downlink latency of every \ac{UE} and reports it to the \ac{CNM}.
A QoS-aware scheduler depending on 5QIs is absent in the OpenAirInterface gNB we used in the \ac{5GS}.
A per-class approach at the gNB scheduler using slicing can be implemented. However, this is optimized only for throughput and not for latency. 
We will work on implementing a custom scheduler that guarantees per-flow latency and reports it to the AF in our future work. 
Hence, for this demo, the worst-case latency of a flow only depends on the rate, burst, and the \ac{TBS} of the UE. 
 
We also added an appendage of the \ac{NW-TT} to the 5G \ac{UPF}.
The \ac{NW-TT} is essentially an extension of the \ac{LCDN} end-host middleware, as it must tag the packets with the correct VLAN and VLAN priority.
However, it cannot be configured to perform the multiple spanning tree protocol used by \ac{LCDN} to partition the network into VLANs for source routing.
Therefore, the \ac{NW-TT} must be configured to tag packets at its egress like other end-hosts using the \code{tc} utility \textit{and} route packets per flow. 
Here, we use the \code{ip route} utility in Linux.
According to 3GPP Release 16, the \ac{NW-TT} can additionally incorporate a hold-and-forward buffer to mitigate the jitter of traffic flows from the 5GS to the DetNet network.
The \ac{TDD} pattern of the 5G system adds significant jitter to the flows as all uplink flows are clustered when the pattern allows for uplink packets and vice-versa. 
Inspired by~\cite{joseifip}, one version of this buffer was implemented to reduce jitter.
The hold-and-forward buffer implements a per-flow or per-class matching by putting the packets in the corresponding queue. 
The mechanism involves retaining the first burst packet for a specified duration, buffering subsequent packets, and releasing them periodically at a pre-configured rate. 

\section{Demonstration}
\label{sec:demo}

The testbed comprises a \ac{5GS} built using simple x-86 PCs and \ac{USRP}.  
The core and \ac{RAN} software stack is OpenAirInterface\cite{openairinterface}, a flexible open-source platform for 5G research.
The core network is built according to the service-based architecture and network function virtualization concept defined by 3GPP for 5G. 
Hence, different network functions are packaged in docker containers.
The \ac{LCDN} \ac{CNM} has a web interface with a Grafana dashboard with three panels. 
The same topology with 2 UEs, one gNB, 3 switches, and 2 end-hosts as shown in Figure \ref{fig:setup} is used for the demonstration. 
In the setup phase (without any UEs connected), the topology discovery manager configures the switches to partition the network into multiple spanning trees, each over a given VLAN ID. 
The switches are also configured to perform \ac{SP} scheduling at the egress. 
For the fixed network topology in the right part of Figure \ref{fig:setup}, three spanning trees can be created. 

\begin{figure}
    \centering
    \includegraphics[width=0.8\columnwidth, height=4cm]{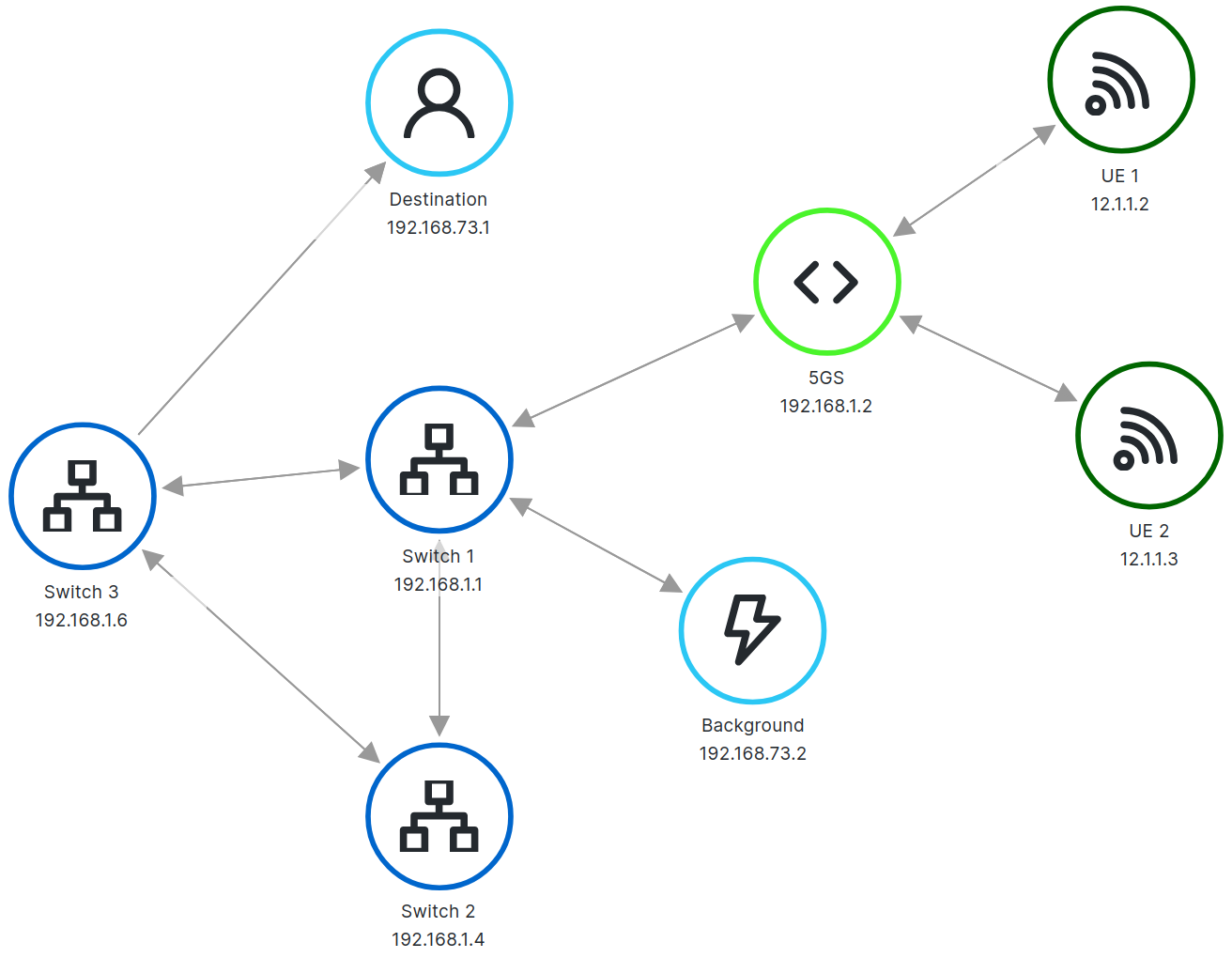}
    \vspace{-0.3cm}
    \caption{\small{Live Topology of the system in Figure 1, as seen by the topology manager at the CNM. While the fixed network is usually static and changes only if a device fails or a cable is unplugged, the 5GS needs a more real-time topology monitoring system due to the mobility of users.}}
    \label{fig:topology_manager_result}
\end{figure}

\begin{figure}
    \centering
    \includegraphics[width=\columnwidth]{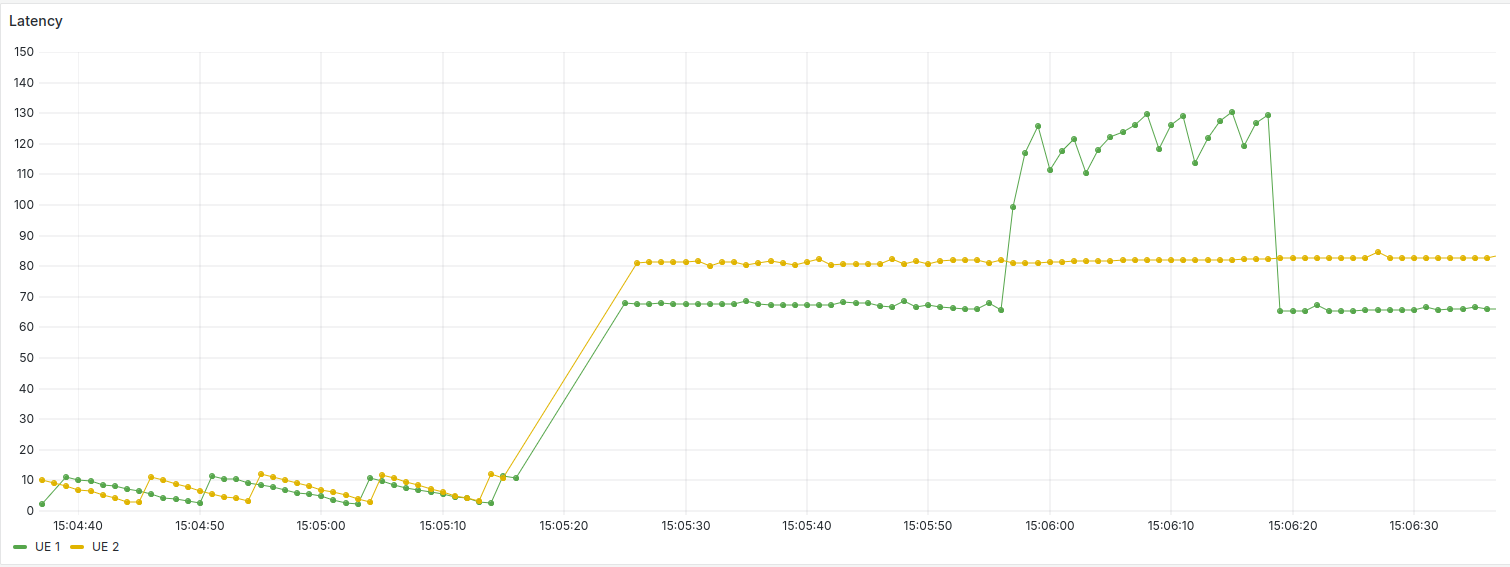}
    \vspace{-0.6cm}
    \caption{\small{The latency of packets from both the UEs to the destination. First, The hold-and-forward buffer at the NW-TT mitigates the jitter caused by the 5G system but increases the overall latency. The latency and packet loss of the critical flow (orange) is preserved even under heavy background traffic.}}
    \label{fig:latency_result}
\end{figure}

The first panel, example in Figure~\ref{fig:topology_manager_result}, renders the network topology built by the topology manager by polling the \code{lldp} information from the switches via a Python-based Telnet client.
When a new UE is connected, it appears on the graph since the CNM knows of a connected UE from the AF.
Due to the dynamic nature of the 5G system, the polling rate of the 5G system by the topology manager is much higher (5s) than that of the fixed network (200s). 
The second panel shows the logs of the \ac{CNM}. 
An entire flow request and installation process can be seen in these logs. 
First, the flow request with TSpec arrives.
Then, the \ac{DNC} pipeline in \ac{LCDN} runs, indicating whether the flow can be accepted. 
If yes, it tells all related end-hosts the VLAN ID and priority to use for that flow and reconfigures any previously installed flows if needed.
It also configures the route and \code{tc} rules at the \ac{NW-TT} via the DetNet \ac{AF}. 
The third panel plots the latency of the packets coming from the two UEs at the destination as shown in Figure \ref{fig:latency_result}.

We use this panel to demonstrate how the latency of a flow changes with changing network conditions and configurations. 
In the demonstration, a flow with a higher priority (orange) is given a higher VLAN priority in class-based scheduling and routed through a less congested path and remains undisturbed as compared to the best effort (green) in the right part of Figure \ref{fig:latency_result}. 
We also use the plots here to demonstrate the effect of the de-jittering function by the hold and forward buffer.
This function increases the overall latency but reduces the jitter in the left part of Figure \ref{fig:latency_result}.
The background traffic and de-jittering function will be turned on periodically and switched on and off during the demonstration. 

\section{Conclusion and Outlook}
\label{sec:conclusion}
In this paper, we present a proof of concept of a unified 5G-DetNet system. 
The components used in the demonstration are low-cost switches, \ac{USRP}, x-86 PCs, and open-source software.
The use of accessible hardware and software solutions not only reduces the overall setup costs but also promotes wider adoption of such an integrated 5G-DetNet testbed for academia and industries.
We demonstrate how our approach can route flows to avoid congestion and meet the specified latency deadline. 
The \ac{5GS} is seen simply as a forwarding device by the \ac{CNM} as specified by the standard in 3GPP Release 16.
The operation of the CNM can be monitored in real-time, via the three panels on a Grafana dashboard. 
In our future work, we plan to work on a flow-based QoS-aware scheduler at the gNB that can provide different forwarding latencies for different flows in the \ac{5GS}. 
\vspace{-0.2cm}

\bibliography{netsoft_demo.bib}
\bibliographystyle{IEEEtran}

\end{document}